\documentclass[letterpaper, 10 pt, conference]{ieeeconf}

\IEEEoverridecommandlockouts                              
\def\isarxiv{1}
\newcommand{\arxiv}[1]{\if\isarxiv1 #1 \fi}

\usepackage[utf8]{inputenc}
\usepackage{graphics} 
\usepackage{xcolor}
\usepackage{graphicx}
\usepackage{caption}

\usepackage{amsmath,amsfonts, amsthm, amssymb}
\usepackage{bbm, bm}
\usepackage{subcaption}
\usepackage{listings}
\usepackage{hyperref}
\usepackage{tcolorbox}
\usepackage{pifont}
\usepackage{tabularx}
\usepackage{algorithmicx}
\usepackage[noend]{algpseudocode}
\usepackage{algorithm}
\usepackage{tablefootnote}
\usepackage{float}
\usepackage{nicematrix}
\usepackage{tikz}
\usepackage{booktabs}
\usepackage{empheq}
\usetikzlibrary{arrows}
\usetikzlibrary{positioning}

\usepackage[usestackEOL]{stackengine}

\newtheorem{theorem}{\textbf{Theorem}}
\newtheorem{lemma}{\textbf{Lemma}}

\newcommand{\UB}{\text{UB}}
\newcommand{\LB}{\text{LB}}
\newcommand{\APXCFO}{\textsf{APXCFO}}
\newcommand{\apxtest}{\textsf{APXTEST}}
\newcommand{\OPT}{\text{OPT}}

\newcommand{\trianglebullet}{$\mbox{\ensuremath{\rhd}}$}

\ifodd 1

    \newcommand{\com}[2]{\textbf{\color{blue} (COMMENT from [#1]: #2)}}
\else

    \newcommand{\com}[2]{}
\fi

\title{\LARGE \bf
    Minimizing Carbon Footprint for Timely E-Truck Transportation: Hardness and Approximation Algorithm
}

\author{
    Junyan Su, Qiulin Lin, Minghua Chen, Haibo Zeng
    \thanks{Junyan Su, Qiulin Lin and Minghua Chen are with City University of Hong Kong, Hong Kong, China (\{junyan.su@my., qiulin.lin@, minghua.chen@\}cityu.edu.hk). Haibo Zeng is with Virginia Tech (hbzeng@vt.edu). }
    \thanks{Corresponding author: Minghua Chen.}
}%

\begin{document}

\maketitle
\thispagestyle{plain}
\pagestyle{plain}

\begin{abstract}
    Carbon footprint optimization (CFO) is important for sustainable heavy-duty e-truck transportation. We consider the CFO problem for timely transportation of e-trucks, where the truck travels from an origin to a destination across a national highway network subject to a deadline. The goal is to minimize the carbon footprint by orchestrating path planning, speed planning, and intermediary charging planning. We first show that it is NP-hard even just to find a feasible CFO solution. We then develop a $(1+\epsilon_F, 1+\epsilon_\beta)$ bi-criteria approximation algorithm that achieves a carbon footprint within a ratio of $(1+\epsilon_F)$ to the minimum with no deadline violation and at most a ratio of $(1+\epsilon_\beta)$ battery capacity violation (for any positive $\epsilon_F$ and $\epsilon_\beta$). Its time complexity is polynomial in the size of the highway network, $1/\epsilon_F$, and $1/\epsilon_\beta$. Such algorithmic results are among the best possible unless P=NP. Simulation results based on real-world traces show that our scheme reduces up to 11\% carbon footprint as compared to baseline alternatives considering only energy consumption but not carbon footprint.

\end{abstract}
\section{Introduction}\label{sec:intro}

Electrifying transportation is a promising and continuing trend toward achieving carbon neutrality for fighting climate change. Within the transportation sector, the trucking industry has garnered increasing attention due to its significant environmental impact. For instance, heavy-duty trucks in the US produce 25\% of the CO2 emissions in the transportation sector~\cite{davisTransportation2021}, despite only accounting for 4\% of the total vehicle population. Furthermore, heavy-duty trucks are responsible for 8.8\% of the total carbon emissions in the US.

Adopting electric trucks (E-Trucks) presents significant potential for greening the heavy-duty truck industry as they produce zero CO2 emissions during operation. However, charging E-trucks with energy generated from carbon-intensive resources (e.g., coal or petroleum) instead of carbon-free ones (e.g., renewable energy) incurs a non-trivial \emph{carbon footprint}.
The average \emph{carbon intensity} (unit: kg CO2 emission/kWh) of coal and petroleum generated electricity is $1.02$ kg/kWh and $0.91$ kg/kWh, respectively~\cite{eia2021monthly}. In contrast, renewable energy has a carbon intensity of zero. Therefore, to fully unlock the environmental benefits of E-trucks, it is crucial to charge them using electricity with low carbon intensity.
Meanwhile, the carbon intensity of the electricity exhibits significant temporal and spatial variations due to fluctuating renewable penetration~\cite{maji2022carboncast}. 
This observation, combined with the time-sensitive nature of truck operations~\cite{ashby1987protecting}, emphasizes the need for joint optimization of 
path planning, speed planning and charging planning for e-trucks to minimize their carbon footprints for sustainability.

In this paper, we study the carbon footprint optimization (CFO) of a heavy-duty e-truck, traveling from an origin to a destination across a national highway network subject to a hard deadline, by exploring the complete design space of path planning, speed planning, and en-route charging planning. 

While existing studies focus on minimizing the energy consumption of conventional internal combustion engine (ICE) trucks~\cite{dengEnergyefficient2017,liuEnergyEfficient2018,xuRide2019}, electric trucks~\cite{su2021Energy, zhang2022optimal} and 
electric vehicles~\cite{liuConstrained2017,celaEnergy2014b,baumModeling2020, fontanaOptimal2013,strehler2017energy}, {only a recent work~\cite{su2023follow} is known for carbon-aware truck operations.}
Note that energy-efficient operation may not necessarily lead to a low carbon footprint~\cite{su2023follow}.
Our simulation based on real-world traces also shows that carbon-aware solutions achieve up to 11\% less carbon footprint than the energy-efficient ones. 
Moreover, the temporal-spatial variation of the carbon intensity results in a non-convex and non-monotone  objective that introduces additional challenges. 
{Indeed, existing discretization technique in~\cite{strehler2017energy}, which relies on the monotonicity of the discretized function, is not directly applicable to the CFO problem.
Our approximation algorithm differentiates from the existing ones, e.g.,~\cite{strehler2017energy}, in that we consider a realistic non-convex objective with both necessary battery and deadline constraints. 
Our work also differentiates from~\cite{su2023follow} in that we present new theoretical hardness results (cf. Theorem.~\ref{thm:feasible:hard}) and an approximation algorithm for the CFO problem.}
We summarize our \textbf{contributions} in the following. 

\trianglebullet\ In Sec.~\ref{sec:formulation}, we present the modeling and problem settings of CFO. We then show that it is NP-hard even just to find a feasible solution to the CFO problem, which directly implies that the NP-hardness of the problem.

\trianglebullet\ In Sec.~\ref{sec:approach_soc}, we develop a bi-criteria (1+$\epsilon_F$, 1+$\epsilon_\beta$)-approximation algorithm that achieves at most $(1+\epsilon_F)$ carbon footprint of the optimal solution with at most $(1+\epsilon_\beta)$ battery capacity, for arbitrary $\epsilon_F$ and $\epsilon_\beta$. The computational complexity of our algorithm is polynomial in the network size, $1/\epsilon_F$, and $1/\epsilon_\beta$. Such algorithmic results are among the best possible unless P=NP. 

\trianglebullet\ In Sec.~\ref{sec:simulation}, we conduct simulations on real-world traces and the results show that our carbon-footprint optimized scheme saves up to 11\% carbon footprint as compared to baseline alternatives considering only energy consumption but not carbon footprint.

\if\isarxiv0
Due to the space limitation, all proofs are included in the technical report~\cite{}.
\fi

\section{Models and Problem Settings}
\label{sec:formulation}

In this section, we discuss the system model and problem formulation of our proposed carbon footprint optimization (CFO) problem. We further show that the problem is NP-hard, even for finding a feasible solution.

\textbf{Transportation Graph and Energy Consumption.}
We consider an e-truck traveling across a national highway system modeled by a directed graph $G=(V,E)$. Here $V=V_r \cup V_c$ is the set of nodes, where $V_r$ denotes the set of nodes connecting highway road segments and $V_c$ denotes the set of nodes for the charging stations. {We denote the set of edges by $E$ 
where each edge represents a road segment.}  
For each edge $e \in E$, we define its minimum and maximum traveling time by $t_e^{lb}$ and $t_e^{ub}$ respectively. 
Without loss of generality, we assume homogeneous road condition (e.g., road grade) on each road segment and define the function $c_e(t_e): [t_e^{lb}, t_e^{ub}] \rightarrow \mathbb{R}$ as the \emph{energy consumption function} for an e-truck traveling across edge $e$ with traveling time $t_e$. 
In practice, $c_e(t_e)$ is non-increasing over $[t_e^{lb}, t_e^{ub}]$ as there is no benefit to drive with both high energy consumption and large traveling time~\cite{dengEnergyefficient2017}. {We denote the minimum and maximum energy consumption for the road segment $e\in E$ as $c_e^{lb},c_e^{ub}$, respectively. 
} {Note that for e-trucks, $c_e(t_e)$ can take negative values over certain road segments (e.g., downhill paths) because of regenerative breaking~\cite{baumShortest2019}.}

\arxiv{
\begin{figure}[!tb]
	\centering
	\includegraphics[width=0.8\linewidth]{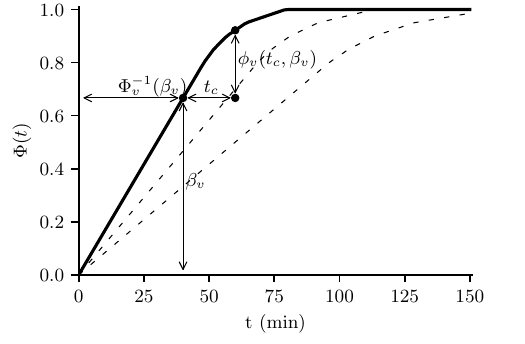}
	\caption{An example of the charging function $\Phi$, and the corresponding SoC difference function $\phi$. The initial SoC is $\beta_v=0.67$. After charging for $20$ minutes, 
  the increment of SoC $\phi(t_{c}, \beta_i)=0.25$. The two dashed lines represent charging functions for different charging schemes with different charging rates.}
	\label{fig:charge}
\end{figure}
}

\textbf{E-Truck Charging.}
We model that an e-truck needs to make two decisions at a charging station: wait for $t_w \in [t_w^{lb}, t_w^{ub}]$ unit of time and charge for $t_c \in [0, t_c^{ub}]$ unit of time. Here we allow an e-truck to wait at the charging station for cleaner electricity. The lower bound $t_w^{lb}$ denotes the time overhead at the charging station (for e.g., parking and paying the bills). 
We model the charging process at charging station $v$ as a concave function $\Phi_v(t)$, which represents the final State-of-Charge (SoC) from zero after charging for $t$ unit of time. We denote by $\beta_v$ as the SoC when the e-truck arrives at node $v \in V_c$. Then, with initial SoC $\beta_v$, the increment of SoC after a charging time $t_c$ at the charging station $v\in V_c$ can be determined as
\begin{equation}
\phi_{v}(t_c, \beta_v) = \Phi_{v}(\Phi^{-1}_{v}(\beta_v) + t_{c}) - \beta_{v}, \label{eq:soc_difference_function}
\end{equation}
where $\Phi_v^{-1}$ is the inverse function of $\Phi_v$. \arxiv{We present a geometric interpretation of $\phi_v$ in Fig.~\ref{fig:charge}.}

\textbf{SoC Evolution.}
According to the energy consumption model, the SoC decreases by $c_e(t_e)$ traveling over edge $e=(u,v) \in E$ with $v \notin V_c$ with time $t_e$, i.e.,
\begin{align}
    \beta_v = \min \left\{B,  \beta_u - c_e(t_e) \right\}, \label{eq:socupdate:energy}
\end{align}
where $B$ is the battery capacity.
According to the energy consumption and the charging model, suppose an e-truck travels across edge $e=(u,v)\in E$ with traveling time $t_e$ and charges at  $v\in V_c$ with charging time $t_c$, the SoC update rule is given by
\begin{align}
    \beta_v = \min \{B, \beta_u - c_e(t_e) + \phi_v(t_c, \beta_u)\}. \label{eq:socupdate:charge}
\end{align}

\textbf{Carbon Intensity and Carbon Footprint.}
For each charging station $v \in V_c$, we model the carbon intensity as a piecewise monotone function $\pi_v(\tau)$ (unit: kg/kWh) with  $\tau$ being the moment of charging process , i.e.,
\begin{equation}
\label{eq:carbon-intensive}
    \pi_v(\tau) = \left\{  
    \begin{aligned}
        & \pi_{v,1}(\tau), &\text{if } \tau \in [0, \tilde{\tau}_1); \\
        & \ldots, & \ldots; \\
        & \pi_{v,n_\pi}(\tau), &\text{if } \tau \in [\tilde{\tau}_{n_\pi -1}, \tilde{\tau}_{n_\pi}];
    \end{aligned} \right.
\end{equation}
{In~\eqref{eq:carbon-intensive}, $ \pi_v(\tau)$ consists of $n_{\pi}$ pieces with breaking points $\tilde{\tau}_1, \tilde{\tau}_2, \cdots \tilde{\tau}_{n_{\pi}}$, and each piece $\pi_{v,i}, \forall i =1,...,n_\pi$, is a monotone function with a well-defined inverse function $\pi_{v,i}^{-1}(\cdot)$.} Note that in practice the carbon intensity is measured and forecast at the hourly scale~\cite{maji2022carboncast} and often interpolated as a piece-wise linear function, which fits into our model.

Given the carbon intensity function $\pi_v(\cdot)$ at a charging station $v\in V_c$, the initial SoC $\beta_v$, the arrival moment $\tau_v$, and the charging time $t_c$, the induced carbon footprint is 

\begin{equation}
    F_{v}(\beta_v, t_{c}, \tau_v) = 
    \frac{1}{\eta} \pi_{v}(\tau_v) \phi(t_c, \beta_v),   \label{eq:carbon-footprint}
\end{equation}
where we denote by $ 0<\eta \leq 1$ the charging efficiency of the battery. 
Note that here {we adopt the first-order approximation of the carbon footprint, i.e., we assume the carbon intensity is fixed during the charging process. This is reasonable as the typical charging time is less than $30$ minutes while the carbon intensity is usually measured hourly~\cite{maji2022carboncast}.}  




\textbf{Carbon Footprint Optimization Problem.}
We consider the \emph{Carbon Footprint Optimization} (CFO) problem of an e-truck traveling across a national highway network from an origin $s\in V$ to a destination $d \in V$ subject to the deadline $T$. {For ease of presentation, we consider that the e-truck starts at moment zero, i.e., $\tau_s=0$.} 
The e-truck equips with an initially fully charged battery with capacity $B$. We denote a path by $\vec{p}=[s=p_1,p_2,\cdots,p_{n_p}=d]\in\mathcal{P}$, where $n_p$ is the number of nodes in path $\vec{p}$ and $\mathcal{P}$ is the set of all paths from $s$ to $d$. We denote the traveling time from $p_i$ to $p_{i+1}$ as $t^r_i$ for all $i\in\{1,2,\cdots,n-1\}$. 
{
Further, we denote by $\mathcal{I}_{p} \subset \left\{ 1,...,n_p \right\}$ the set of index for charging stops, i.e., for any $j \in \mathcal{I}_{p}$, $p_j \in V_c$ is a charging stop.}
We define the \emph{Carbon Footprint Optimization} (CFO) problem as follows:
{
\begingroup
\allowdisplaybreaks 
\begin{subequations}
\begin{align}
 \min\; & \sum_{j \in \mathcal{I}_{p}}^{} F_{p_{j}}(\beta_{j},t^c_{j}, \tau_{j})\label{eq:obj}\\
   \text{s.t.}\;&  \sum_{i=1}^{n_p} t^r_i +\sum_{j \in \mathcal{I}_{p}}^{} \left(t^w_{j}+t^c_{j}\right) \leq T,\label{eq:deadline}\\
   & \tau_{j}= \sum_{i=1}^{j-1} t^r_i +\sum_{k \in \mathcal{I}_{p}; k < j}^{} \left(t^w_{k}+t^c_{k}\right) + t^w_{j},\label{eq:arrival}\\
   & \beta_j = \min\Big\{B, \beta_{j-1}-c_{(p_{j-1},p_j)}(t^r_{j-1}) \nonumber\\
   & \;\;\;\;\;\;\;+ \phi_{p_j} \left(t^c_{j},\beta_{j-1}\right) \Big\} , \forall j\in \mathcal{I}_{p} ,\label{eq:soc_edge}\\
   & \beta_i = \min\{B, \beta_{i-1}-c_{(p_{i-1},p_i)}(t^r_{i-1})\},
   \nonumber\\
   & \;\;\;\;\;\;\;\;\;\;\;\;\;\forall i\in\{1,2,\cdots,n_p\} \backslash \mathcal{I}_{p} \label{eq:soc_station}\\
   & \beta_i\geq 0,\forall i\in\{1,2,\cdots,n_p\},\label{eq:soc_postive}\\
   & \beta_0=B,\label{eq:soc_initial}\\
   \text{var.}\; & \vec{p}\in\mathcal{P},\\
   & t^r_i\in [t^{lb}_{(p_{i-1},p_i)},t^{ub}_{(p_{i-1},p_i)}],  \beta_i,\forall i\in\{1,2,\cdots,n_p\}, \\
   &t^w_{j}\in[t^{lb}_w,t^{ub}_w],t^c_{j}\geq 0,\tau_j,\forall j \in \mathcal{I}_{p}.
\end{align}
\end{subequations}
\endgroup
}


Our objective in ~\eqref{eq:obj} is the carbon footprint incurred by charging at $m_p = |\mathcal{I}_{p}|$ charging stops. The deadline constraint is in~\eqref{eq:deadline}. In~\eqref{eq:arrival}, we obtain the arrival time at $j$-th stop, which is required for determining the carbon footprint in~\eqref{eq:obj}. The evolution of SoC is determined by~\eqref{eq:soc_edge} and~\eqref{eq:soc_station}, according to our discussions in~\eqref{eq:socupdate:energy} and~\eqref{eq:socupdate:charge}. We require the SoC to be always non-negative in~\eqref{eq:soc_postive} and set the initial SoC as the battery capacity in~\eqref{eq:soc_initial}.

\begin{theorem}
\label{thm:feasible:hard}
       The CFO problem is NP-hard. Further, it is NP-hard even just to find a feasible solution for the problem.
\end{theorem}
\arxiv{We leave the proof to Appendix~\ref{sec:thm:hard}.}
Theorem~\ref{thm:feasible:hard} implies that unless P=NP, one could not expect to solve the problem or find a feasible solution within a time complexity polynomial to the network size and the input size of $B$ and $T$. {In such case, the best possible algorithmic results one could expect is then a polynomial time bi-criteria approximation scheme~\cite{horvath2018multi}.} We will design one such algorithm and characterize its performance in the next section. 



\section{A Bi-criteria Approximation Algorithm}\label{sec:approach_soc}
In this section, we develop a $(1+\epsilon_F, 1+\epsilon_\beta)$ bi-criteria approximation algorithm that achieves a carbon footprint within a ratio of $(1+\epsilon_F)$ to the minimum with no deadline violation and with at most a ratio of $(1+\epsilon_\beta)$ battery capacity violation (for any positive $\epsilon_F$ and $\epsilon_\beta$) and a time complexity polynomial in the size of the highway network, $1/\epsilon_F$, and $1/\epsilon_\beta$.
The essence of our approximation algorithm is a test procedure as outlined in Algorithm~\ref{alg:apxtest}. 
Given a guess $\Omega>0$ of the optimal objective $\OPT$ to the CFO problem, Algorithm~\ref{alg:apxtest} approximately compare them and report if $\OPT > \Omega $ or $\OPT\leq(1+\epsilon_F)\Omega$. 
Then we use a bisection technique (cf. Algorithm~\ref{alg:apx:cfo}) over the possible range of $\OPT$ with this test procedure to obtain an $\Omega$ close enough to $\OPT$ and recover its corresponding solution to the CFO problem.

In Algorithm~\ref{alg:apxtest}, we seek to find a feasible path that is within the given objective upper bound $\Omega$ or report infeasibility otherwise. 
We discretize the SoC and the objective value and then use dynamic programming (DP) to obtain the optimal path under discretized values. 
We carefully determine the discretization precision to achieve the $(1+\epsilon_F, 1+\epsilon_\beta)$ approximation while ensuring polynomial time complexity. 
Meanwhile, our discretization procedure guarantees that DP is able to find a feasible solution if $\OPT \leq (1+\epsilon) \Omega$.

We divide the algorithm into two phases.
In the first phase, we compute the energy-constrained fastest path from a charging station to other nodes without intermediary charging stops. In the second phase, we decide the charging locations and scheduling by utilizing the information from phase I. Then we recover the solution based on the results of those two phases or report infeasible otherwise.

\begin{algorithm}[!tb]

    \caption{$\apxtest(s, d, T, B, \Omega, \epsilon_\beta, \epsilon_F)$ 
    }\label{alg:apxtest}
    \begin{algorithmic}[1]
        \State $\delta_\beta \gets \frac{\epsilon_\beta B}{|V|+|V_c|+1}$, 
        ${M}_{\beta} \gets \lceil \frac{|V|+|V_c|+1}{\epsilon_\beta} \rceil + |V| + |V_c| + 1$ \label{alg:apxtest:deltab}
        \State $\delta_F \gets \frac{\epsilon_F \Omega}{|V|+|V_c|+1}$, $M_F \gets \lceil \frac{|V|+|V_c|+1}{\epsilon_F}\rceil + |V| + |V_c| + 1 $ \label{alg:apxtest:deltaF}
        \State $\hat{\beta_0} \gets \lfloor (1+\epsilon_\beta) B / \delta_\beta \rfloor \cdot \delta_\beta $

        \State \textbf{Phase I: energy-constrained path without charging stops.} \label{alg:apxtest:phase1}
        \For{$\forall i_u \gets 0,1...,M_\beta, \forall u \in V_c \cup \{s\}$}
            \State $\beta_u \gets i_u \delta_\beta$
            \State $\psi_{u,\beta_u}(u, \beta_u) \gets 0$ and $\psi_{u,\beta_u}(\cdot, \cdot) \gets \infty$ otherwise.
            \For{$\forall i_v\gets 0,...,M_\beta, \forall v \in V$}
                \For{$\forall (w,v) \in E, \forall i_c \gets \lfloor \frac{c_e^{lb}}{\delta_\beta} \rfloor ... \lfloor \frac{c_e^{ub}}{\delta_\beta} \rfloor $}
                    \State $(\beta_v,\, \hat{c}) \gets (i_v\delta_\beta, \, i_c\delta_\beta)$
                    \State $\hat{\beta}_w \gets \beta_w(\beta_v, \hat{c})$ according to \eqref{eq:psi:betaw}.
                    \If{$\beta_w \in [0, (1+\epsilon_\beta) B]$}
                        \State Update $\psi_{u,\beta_u}(v,\beta_v)$ according to \eqref{eq:psi:update}. \label{alg:apxtest:psiupdate} 
                    \EndIf
                \EndFor
            \EndFor
        \EndFor \label{alg:apxtest:phase1end}

        \State \textbf{Phase II: charging planning.} \label{alg:apxtest:phase2}
        \State $\sigma_{s,\hat{\beta}_0}(s,0,\hat{\beta}_0) \gets 0$ and $\sigma_{s,\hat{\beta}_0}(\cdot,\cdot,\cdot) \gets \infty$ otherwise.
            \label{alg:apxtest:sigmainit}

        \For{$\forall i_\rho \gets 0,...,M_F$,\, $\forall i_F \gets 0,...,i_\rho$} \label{alg:apxtest:rhoF}
        \For{$\forall i_c \gets 0,...,M_\beta$,
             $\forall i_v \gets 0,...,M_\beta$,
             $\forall i_u \gets 0,...,M_\beta$, \label{alg:apxtest:sigmab}
        }
 
        \For{$\forall u\in V_c \cup \{s\},\, \forall v \in V_c \cup \{d\}$}
            \State $(\rho, \hat{F}) \gets (i_\rho \delta_F, i_F \delta_F)$
            \State $(\beta_u, \beta_v, \beta_c) \gets (i_u \delta_\beta, i_v \delta_\beta, i_c \delta_\beta)$

            \State $\beta_u' = \min\{B, \beta_{u} + \beta_c\}$

            \State $\sigma_u \gets \sigma_{s,\hat{\beta}_0}(u, \rho - \hat{F}, \beta_{u}) $
            \State $t_w \gets g_v(\beta_{u}, t_c, \sigma_u, \hat{F} )$ (cf. problem~\eqref{prob:g:tw}) \label{alg:apxtest:tw}


            \State \Longunderstack{$t_c \gets \Phi_u^{-1}( \beta_u' ) - \Phi_u^{-1}(\beta_{u})$} \label{alg:apxtest:tc}
            \State $t_r \gets \psi_{u,\beta_u'}(v, \beta_v)$ \label{alg:apxtest:tr}

            \State $\tilde{\sigma} \gets \sigma_u + t_r + t_w + t_c $
            \State $\sigma_v \gets \sigma_{s,\hat{\beta}_0}(v, \rho, \beta_v) $
            \If{ $\tilde{\sigma} \leq \sigma_v$ }
                \State $\sigma_{s,\hat{\beta}_0}(v, \rho, \beta_v) \gets \tilde{\sigma}$ \label{alg:apxtest:sigmaupdate}
            \EndIf
        \EndFor
        \EndFor
        \EndFor \label{alg:apxtest:phase2end}

        \State Let $i^*_\rho \in \left\{ 0,..., M_F \right\}$ be the minimal $i_\rho$ such that $\sigma_{s,\hat{\beta}}(d, i_\rho \delta_F,  0 ) \leq T $
        \If{$\rho^*$ exists}
            \State sol $\gets$ the solution corresponding to $\rho^*$ \label{alg:apxtest:return}
        \Else
            \State sol $\gets$ NULL \label{alg:apxtest:null}
        \EndIf
        \State \Return sol
    \end{algorithmic}
\end{algorithm}

\textbf{Phase I: Energy-Constrained Fastest Path from Single Source.} 
We use the rounding and scaling technique~\cite{dengEnergyefficient2017, strehler2017energy} to ensure the performance and guarantee polynomial time complexity. 
In particular, we discretize the SoC into $M_\beta$ states with precision $\delta_\beta$ (cf. line~\ref{alg:apxtest:deltab} of Algorithm~\ref{alg:apxtest}). 
For each discretized energy consumption $\hat{c}$, we compute its corresponding travel time by
\begin{equation}
    \hat{t}_e(\hat{c}, \delta_\beta) = \left\{  
        \begin{aligned}
            & t_e^{lb},& \text{if } \lceil \frac{\hat{c}}{\delta_\beta} \rceil = \lceil \frac{c_e^{ub}}{\delta_\beta} \rceil; \\
            & c_e^{-1}(\hat{c}), & \text{if } \hat{c} \in [c_e^{lb}, c_e^{ub});  \\
            & \infty, & \text{otherwise}.
        \end{aligned}
    \right. \label{eq:t:hat}
\end{equation}
Recall that $c_e^{lb},c_e^{ub}$ are the minimum and maximum energy consumption for the road segment $e\in E$, respectively. 

We then use dynamic programming (DP) to compute the energy-constrained fastest solutions without charging stops. 
We denote by $\psi_{u,\beta_u}(v, \beta_v)$ the minimal traveling time from $u$ to $v$ with initial SoC $\beta_u$ and the final SoC lower bounded by $\beta_v$. 
Then the Bellman equation is given by
\begin{align}
        & \psi_{u,\beta_u}(v,\beta_v) =  \min \Bigg\{  \psi_{u,\beta_u}(v,\beta_v-\delta_\beta), 
         \nonumber \\ 
         & \min_{ (w,\hat{c}) \in \mathcal{N}(v,\beta_v) } 
         \Big\{   \psi_{u,\beta_u}\Big(w,\beta_w\left(\beta_v, \hat{c}\right)\Big) + \hat{t}_{(w,v)}(\hat{c}, \delta_\beta) \Big\} \Bigg\},    
\end{align}
with the initial condition $\psi_{u,\beta_u}(u, \beta_u) = 0$. 
Here the previous SoC state $\beta_w$ is given by 
\begin{align}
    \beta_w(\beta_v, \hat{c}) = \left\{ \begin{aligned}
        &(1+\epsilon_\beta) B - \hat{c}, &\text{if } \beta_v = (1+\epsilon_\beta) B \text{ \& } \hat{c} < 0;  \\
        &\beta_v + \hat{c}, &\text{if } (\beta_v + \hat{c}) \in [0, (1+\epsilon_\beta) B]; \\
        & \infty, &  \text{otherwise} .
    \end{aligned} \right. \label{eq:psi:betaw}
\end{align}
Note that here we drop the harvested energy when battery overflow happens.
The state set $\mathcal{N}(v,\beta_v)$ is given by
\begin{align}
    \mathcal{N}(v, \beta_v) = \Bigg\{&  
        (w, \hat{c}) \;:\; e = (w,v) \in E, \nonumber \\
        &\hat{c} \in \Big\{ i_c \delta_\beta : i_c \in \lfloor {c_e^{lb}}/{\delta_\beta} \rfloor , ..., \lfloor {c_e^{ub}}/{\delta_\beta} \rfloor \Big\}, \nonumber \\
        &\beta_w(\beta_v, \hat{c}) \in [0, (1+\epsilon_\beta) B]
        \Bigg\}.
\end{align}
Therefore, given any $u,v \in V$, any discretized states $\beta_u, \beta_v$, any road segment $e=(w,v) \in E$, and discretized energy consumption $\hat{c}$, we update $\psi$ in line~\ref{alg:apxtest:psiupdate} of Algorithm~\ref{alg:apxtest} as follows:
\begin{equation}
    \begin{aligned}
        \psi_{u,\beta_u}(v,\beta_v)  = \min \Big\{ & \psi_{u,\beta_u}(v, \beta_v), \\
         & \psi_{u,\beta_u}(w,\beta_w) + \hat{t}_e(\hat{c}, \delta_\beta) \Big\}.
    \end{aligned} \label{eq:psi:update}
\end{equation}
The overall phase I procedure is summarized in line~\ref{alg:apxtest:phase1}-\ref{alg:apxtest:phase1end} of Algorithm~\ref{alg:apxtest}. For any source node $u$ with the initial SoC $\beta_u$, we update $\psi$ in a bottom-up manner: we first update for all nodes with final SoC $\beta_v = \delta_\beta$, and then $\beta_v = 2\delta_\beta$, until $\beta_v = M_\beta \delta_\beta$ and get the complete value function $\psi$.

\textbf{Phase II: Routing and Planning for Charging.}
In phase II, we discretize the objective value with precision $\delta_F$ and the total number of discretized values $M_F$. We then use DP to compute the approximate solution.
We denote by $\sigma_{s,\hat{\beta}_0}(v, \rho, \beta_v)$ the minimal travel time from the source $s$ to $v \in V_c \cup \{d\}$ with initial SoC $\hat{\beta}_0$, the final SoC lower bound $\beta_v$ and the objective upper bound $\rho$.
Then the Bellman equation is given by
\begin{align}
    \sigma_{s,\hat{\beta}_0}(v, \rho, & \beta_v) = \min \Big\{
    \sigma_{s, \hat{\beta}_0}(v, \rho - \delta_F, \beta_v), 
    \nonumber \\ 
    \min_{(u, t_w, t_c, \beta_u, \hat{F}) \in \mathcal{M}} 
     & \Big[
        \sigma_{s,\hat{\beta}_0}(u, \rho - \hat{F}, \beta_u)  + t_w + t_c
    \nonumber \\ &  + \psi_{u, (\phi_u(t_c, \beta_u) + \beta_u)}\Big( v, \beta_v\Big) 
       \Big]
    \Big\}, \label{eq:sigma:bellman}
\end{align} 
with the initial condition is $\sigma_{s,\hat{\beta_0}}(s, 0, \hat{\beta}_0) = 0$.
The state set $\mathcal{M}$ is given by
\begin{align}
        \mathcal{M} = \Bigg\{ & (u, t_w, t_c, \beta_u, \hat{F}) : \forall u \in V_c \cup \{s\}, 
        \nonumber\\ & F_u(\beta_u, t_c, \tau_u + t_w ) \leq \hat{F}, 
        \nonumber\\ & \tau_u = \sigma_{s,\hat{\beta}_0}( u, \rho-\hat{F}, \beta_u), t_w \in [t_w^{lb}, t_w^{ub}],
        \nonumber\\ & \forall \hat{F}\in \{0,\delta_F,\ldots,\delta_F M_F\}, 
        \nonumber\\ & \forall \beta_u \in \{ 0,\delta_\beta,\ldots,\delta_\beta M_\beta \}, 
        \nonumber\\ & \forall \phi_u(t_c, \beta_u) \in \{ 0,\delta_\beta,\ldots,\delta_\beta M_\beta \}
            \Bigg\}.
\end{align}
Therefore, in phase II, given the discretized objective value $\hat{F}$, the discretized SoC $\beta_u$ at $u$, the SoC $\beta_v$ at $v$, and the charged energy $\beta_c$, we can determine the charging time and the travel time between $(u,v)$ (cf. line~\ref{alg:apxtest:tc} and line~\ref{alg:apxtest:tr} in Algorithm~\ref{alg:apxtest}). 
However, to solve the Bellman equation~\eqref{eq:sigma:bellman}, a simple enumeration over the discretized objective values and battery states is not sufficient 
because the decision variable waiting time $t_w$ remains undetermined. 
To obtain the waiting time that satisfies the Bellman equation~\eqref{eq:sigma:bellman}, we need to find the minimal waiting time subject to the objective upper bound:
\begin{subequations}
    \begin{align}
        g_u(\beta_u, t_c, \tau_u, \hat{F}) = &\min_{t_w \in [t_w^{lb}, t_w^{ub}]}\; t_w  \\
        & \text{s.t. }\;    F_u(\beta_u, t_c, \tau_u + t_w) \leq \hat{F}. \label{cons:g:tw:F}
    \end{align} \label{prob:g:tw}
\end{subequations}
We set $g_v = \infty$ if the problem~\eqref{prob:g:tw} is infeasible. The problem~\eqref{prob:g:tw} is non-convex due to the constraint~\eqref{cons:g:tw:F}. 
However, we are able to solve it by exploring the piecewise monotone property of the carbon intensity function.

\begin{lemma}
    The problem~\eqref{prob:g:tw} can be solved in polynomial time  by finding the minimum waiting times satisfying ~\eqref{cons:g:tw:F} among all pieces in $\pi_u(\tau)$, i.e.
    \begin{equation}
        t^*_w = \min_{i=1,...,n_\pi} \pi_{u,i}^{-1}(\hat{F} / \phi_u(t_c, \beta_u))
    \end{equation}
\end{lemma}


The overall phase II procedure is summarized in line~\ref{alg:apxtest:phase2}-\ref{alg:apxtest:phase2end} of Algorithm~\ref{alg:apxtest}. After initializing the boundary case for $\sigma$, we use DP to update the information of $\sigma$. 
In line~\ref{alg:apxtest:rhoF}, we define the objective upper bound $\rho$ from source $s$ to the node $v$ and the objective upper bound $\hat{F}$ from the node $u$ to $v$.
In line~\ref{alg:apxtest:sigmab}, we denote by $\beta_u$ the arrival SoC at the node $u$, by $\beta_v$ the arrival SoC at the node $v$, and by $\beta_c$ the amount of charged energy at node $u$. 
We then compute the waiting time $t_w$ in line~\ref{alg:apxtest:tw}, the charging time $t_c$ in line~\ref{alg:apxtest:tc}, and the travel time $t_r$ between the node $u$ and the node $v$ in line~\ref{alg:apxtest:tr}.
We then update the $\sigma$ in line~\ref{alg:apxtest:sigmaupdate} according to the Bellman equation.

Overall, the test procedure Algorithm~\ref{alg:apxtest} approximately compare $\Omega > 0$ with the optimal carbon footprint \OPT, and report if $\OPT > \Omega$ or $\OPT \leq (1+\epsilon_F) \Omega$ as summarized in the following lemma.




\begin{lemma} \label{lem:apxtest:result}
    If the CFO problem is feasible and $\Omega \geq \OPT$, then Algorithm~\ref{alg:apxtest} must return a solution profile $(\vec{p}, \vec{t}^r, \vec{t}^w, \vec{t}^c)$ with the SoC profile $\vec{\beta}$ and the initial SoC $\hat{\beta}_0=(1+\epsilon_\beta)B$ that satisfy the following conditions:
    \vspace{-1mm}
    \makeatletter
    \@fleqntrue
    \makeatother
    \begin{subequations}
        \begin{flalign}
           a.\quad & \sum_{i=1}^{n_p} t^r_i + \sum_{j \in \mathcal{I}_{p}}^{} (t^w_j + t^c_j) \leq T, \label{lem:apxtest:ineq:ddl}\\
            b.\quad & 0  \leq \beta_i  \leq (1+\epsilon_\beta) B, \label{lem:apxtest:ineq:b} \\
            c.\quad & F(\vec{p}, \vec{t}^r, \vec{t}^w, \vec{t}^{c})  \leq \OPT + \epsilon_F \Omega. \label{lem:apxtest:ineq:obj}
        \end{flalign} \label{lem:apxtest:ineq}
    \end{subequations}
    \vspace{-6mm}
\end{lemma}

\arxiv{We leave the proof details to Appendix~\ref{sec:lem:apxtest}.}
The proof idea is to round the optimal solution to the discretized solution such that the rounded solution satisfies the conditions~\eqref{lem:apxtest:ineq} and we can ensure that the rounded solution has been examined in the loops of Algorithm~\ref{alg:apxtest}. 

Lemma~\ref{lem:apxtest:result} implies that if Algorithm~\ref{alg:apxtest} returns a feasible solution in line~\ref{alg:apxtest:return}, then we have $\OPT \leq (1+\epsilon_F) \Omega$ and if Algorithm~\ref{alg:apxtest} returns an empty solution in line~\ref{alg:apxtest:null}, we have $\OPT \geq \Omega$. This result naturally leads to our main algorithm which bisects over the possible range of $\OPT$.


\textbf{Main Algorithm and Performance Analysis.}
We now present the main algorithm outlined in Algorithm~\ref{alg:apx:cfo}. 
We first set the lower and upper bound of the optimal objective value in lines~\ref{alg:apx:cfo:1}-\ref{alg:apx:cfo:2}.
At each iteration we maintain a lower bound and upper bound for the optimal objective value \OPT\ and update one of the bounds after probing the middle point $\Omega$. We terminate the iteration and recover the approximate solution when the ratio between the upper bound and the lower bound is less than two.
We summarize the main results for Algorithm~\ref{alg:apx:cfo} in the following two theorems.

\begin{algorithm}[!tb]

    \caption{\APXCFO$(s, d, T, \epsilon_\beta, \epsilon_F)$}\label{alg:apx:cfo}
    \begin{algorithmic}[1]
        \State $\LB \gets 0$ \label{alg:apx:cfo:1}
        \State $\UB \gets |V_c| B \pi^{ub} $ \label{alg:apx:cfo:2}
        \While{$\frac{\UB}{\LB} > 2$}
            \State $\Omega \gets (\LB+\UB)/2$
            \If{\apxtest$(s,d,T,\Omega,\epsilon_\beta, \epsilon_F/2)=$ NULL}
                \State $LB \gets \Omega$
            \Else
                \State $UB \gets \Omega$
            \EndIf
        \EndWhile
        \State sol $\gets$ \apxtest$(s,d,T,\UB,\epsilon_\beta, \epsilon_F/2)$
        \State \Return sol
        
    \end{algorithmic}
\end{algorithm}


\begin{theorem}\label{thm:apx:result}
    If the CFO problem is feasible, then Algorithm~\ref{alg:apx:cfo} must return a solution profile $(\vec{p}, \vec{t}^r, \vec{t}^w, \vec{t}^c)$ with the initial SoC $\hat{\beta}_0 = (1+\epsilon_\beta)B$ and the SoC profile $\vec{\beta}$ that satisfy~\eqref{lem:apxtest:ineq:ddl},~\eqref{lem:apxtest:ineq:b} and 
    \begin{equation}
        F(\vec{p}, \vec{t}^r, \vec{t}^w, \vec{t}^{c})  \leq (1 + \epsilon_F) \OPT. \label{thm:apx:result:obj}
    \end{equation}
\end{theorem}
\arxiv{
\begin{proof}
    The inequalities~\eqref{lem:apxtest:ineq:ddl} and~\eqref{lem:apxtest:ineq:b} directly follows from Lemma~\ref{lem:apxtest:result}. To bound the objective, we have
    \begin{align}
        F(\vec{p}, \vec{t}^r, \vec{t}^w, \vec{t}^{c}) & \leq \OPT + (\epsilon_F/2) \Omega 
         \leq \OPT + (\epsilon_F/2) \UB \nonumber
        \\ & \leq \OPT + \epsilon_F \LB 
        \leq (1+\epsilon_F) \OPT, 
    \end{align}
    which completes our proof.
\end{proof}
}


%
%

\begin{theorem} \label{thm:apx:complexity}
    The time complexity of Algorithm~\ref{alg:apx:cfo} is 
    $O\left( \left( 1+ 1/\epsilon \right)^5 n^5 |V_c|^2 n_\pi \log \UB  \right)$,
    where $n=|V|+|V_c|$ and $\epsilon=\min\{\epsilon_\beta, \epsilon_F \}$. 
\end{theorem}

By combining Theorem~\ref{thm:apx:result} and Theorem~\ref{thm:apx:complexity}, we show that Algorithm~\ref{alg:apx:cfo} is a $(1+\epsilon_F, 1+\epsilon_\beta)$ bi-criteria approximation scheme that achieves a carbon footprint within a ratio of $(1+\epsilon_F)$ to the optimal carbon footprint with no deadline violation and at most a ratio of $(1+\epsilon_\beta)$ battery violation for any positive $\epsilon_F$ and $\epsilon_\beta$. The time complexity of Algorithm~\ref{alg:apx:cfo} is polynomial in the network size, $1/\epsilon_F$, $1/\epsilon_\beta$. The algorithm is among the best possible given the hardness of the CFO problem (cf. Theorem~\ref{thm:feasible:hard}).

\textbf{Discussion on constraint violation.} Our proposed approximation algorithm may result in battery capacity violations (cf. \eqref{lem:apxtest:ineq:b}). 
To obtain a feasible solution without violating the battery capacity constraints of size $B$, we can use \APXCFO\ with a capacity of $B/(1+\epsilon_\beta)$. We refer to the CFO instance with a capacity of $B$ as $\textsf{CFO}(B)$. If $\textsf{CFO}(B/(1+\epsilon_\beta))$ with a reduced battery capacity of $B/(1+\epsilon_\beta)$ is also feasible\footnote{Note that the CFO problem with a smaller battery capacity may not be feasible, since the feasible region shrinks and may be empty.}, then $\APXCFO$ can generate a strictly feasible solution with a capacity of $B$ and a deadline of $T$.  In practice, it should be easy for $\textsf{CFO}(B/(1+\epsilon_\beta))$ to be feasible by choosing a small enough $\epsilon_\beta > 0$.


\section{Numerical Experiment}\label{sec:simulation}


\subsection{Experimental Setup}

\emph{Transportation network.} 
We collect the highway network data from the Map-based Educational Tools for Algorithm Learning (METAL) project~\cite{metalproject}, the road speed data from HERE Maps, and the elevation data from the Shuttle Radar Topography Mission (SRTM) project. We consider a typical origin-destination pair from Atlanta to Nashville, representing 7.5 billion dollars of transported freight~\cite{hwang2016freight}.
We pre-process the network to the relevant region and merge the edges with the same grade. The reduced network has $48$ nodes and $242$ edges.

\emph{Charging station data.} We collect the charging station locations from the OpenStreetMap (OSM) and obtain $11$ charging stations in the considered region. We consider a piecewise linear charging function~\cite{baumShortest2019}\arxiv{as illustrated in Fig.~\ref{fig:charge}}. We then collect and compute the carbon intensity data from Energy Information Administration (EIA)~\cite{eia2021monthly}.

\emph{Energy consumption model.} We consider an e-truck with battery capacity $B=300$ kWh with total weight of $36$ tons. We use the simulator \textsf{FASTSim}~\cite{gonderFuture2018} to collect the energy consumption data under different speeds and grades and then fit the data with cubic polynomial functions.


\emph{Algorithm Comparison.} We implement and compare the following approaches.

\trianglebullet\ \textsf{APX-C}: our approximation algorithm with carbon footprint objective. We set $\epsilon_\beta = \epsilon_F = 0.1$.

\trianglebullet\ \textsf{APX-E}: the energy-efficient baseline. We use our approximation algorithm with the energy minimization objective by setting the carbon intensity $\pi(\cdot) \equiv 1 $. We also set $\epsilon_\beta = \epsilon_F = 0.1$.

\trianglebullet\ \textsf{LB}: the lower bound of the optimal objective provided by Theorem~\ref{thm:apx:result}. 
\subsection{Benefit of Carbon Footprint Optimization}

\begin{figure}[!tb]
    \centering
    \includegraphics[width=.9\linewidth]{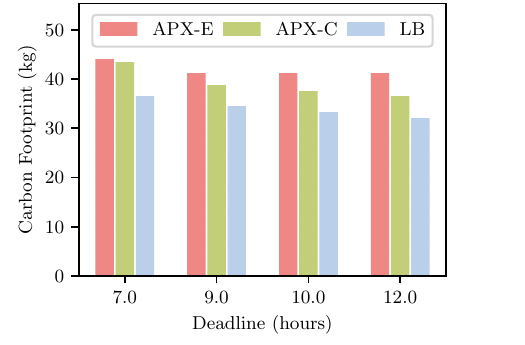}
    \vspace{-2mm}
    \caption{Carbon footprint of different baselines with respect to the deadline.}\label{fig:apx:ddl}
    \vspace{-5mm}
\end{figure}

We vary the deadline from $7$ hours to $12$ hours and present the result in Fig.~\ref{fig:apx:ddl}. We observe that when the deadline is tight (e.g., $7$ hours), \textsf{APX-E} and \textsf{APX-C} incur similar carbon footprint. Meanwhile, as the deadline gets relaxed, the solutions of both \textsf{APX-E} and \textsf{APX-C} incur less carbon footprint, but our scheme \textsf{APX-C} gives larger carbon footprint reduction and its performance is closer to the lower bound. 
In particular, when the deadline is set to 12 hours, our proposed \textsf{APX-C} algorithm can achieve an 11\% reduction in carbon footprint compared to \textsf{APX-E}. These findings highlight that an energy-efficient solution, such as that produced by \textsf{APX-E}, may not necessarily optimize carbon footprint since it does not consider the carbon intensity during charging. It is therefore imperative to implement the intelligent carbon footprint aware design, which can result in substantial carbon reduction as shown in Fig.~\ref{fig:apx:ddl}.



\section{Concluding Remarks}

We address the problem of optimizing carbon footprint (CFO) for sustainable and timely transportation of E-trucks. We prove that even finding a feasible solution for this problem is NP-hard. We then develop a $(1+\epsilon_F, 1+\epsilon_\beta)$ bi-criteria approximation algorithm that achieves a carbon footprint within a ratio of $(1+\epsilon_F)$ to the minimum with no deadline violation and at most a ratio of $(1+\epsilon_\beta)$ battery capacity violation. 
Through simulations, we show that our algorithm can achieve up to an 11\% reduction in carbon footprint compared to baseline alternatives. 
We remark that the method used to design our algorithm in this paper can also be applied to develop a distinct approximation algorithm that avoids violating battery capacity but may violate deadlines. Further, the formulation and approach can be extended to address other problems, such as minimizing fuel costs for ICE-truck transportation that requires gas refills under geographically and temporally varying gas prices.



\section*{Acknowledgement}
The authors would like to thank the anonymous reviewers for their insightful comments that help improve this work. 
The work presented in this paper was supported in part by a General Research Fund from Research Grants Council, Hong Kong (Project No. 14207520).

\bibliographystyle{IEEEtran}
\bibliography{ref}

\arxiv{
\appendix
\section{Appendix}

\subsection{Proof of Theorem~\ref{thm:feasible:hard}} \label{sec:thm:hard}

\begin{proof}
    {We prove that it is NP-hard to find a feasible solution for CFO. And it directly implies that CFO is NP-hard.} We prove it by reducing from the NP-hard PASO problem~\cite{dengEnergyefficient2017} to the problem of finding a feasible solution for the CFO problem. {The PASO problem consider minimizing the energy consumption of an ICE-truck (no need to consider charging) by optimizing the speeding planning and path planning with a hard deadline constraint.} 

    Suppose we have a polynomial algorithm $\mathcal{A}$ for finding a feasible solution for the CFO problem with time complexity $O(h)$. 
    For any PASO problem with deadline $T$, we consider an instance of CFO with the following parameters: (i) the transportation graph and the deadline are the same as PASO; (ii) there is no charging stations in the graph; (iii) the e-truck is fully charged with capacity $B$ at the origin.

    We denote such instance with capacity $B$ as CFO($B$). 
    By construction, if CFO($B$) is feasible, then $B$ is an objective upper bound for the PASO problem. On the other hand, if CFO($B$) is infeasible, then $B$ is an objective lower bound for the PASO problem.
    Since the algorithm $\mathcal{A}$ must return a feasible solution whenever CFO($B$) is feasible, we can apply a bisection technique over the optimal objective of PASO and find the optimal solutions to PASO with complexity $O(\log \text{UB})$ where UB is an upper bound of the optimal value for the PASO.
    Therefore, we give a reduction from the PASO problem to the problem of finding a feasible solution for the CFO problem.
\end{proof}

\subsection{Proof of Lemma~\ref{lem:apxtest:result}} \label{sec:lem:apxtest}

\begin{proof}
    \newcommand{\roundsol}{(\vec{\hat{p}}, \vec{\hat{t}}^{r}, \vec{\hat{t}}^{w}, \vec{\hat{t}}^{c})}
    If the CFO problem is feasible, we denote by $(\vec{p}^*, \vec{t}^{r*}, \vec{t}^{w*}, \vec{t}^{c*})$ the optimal solution profile with SoC profile $\vec{\beta}^{*}$ the objective $\OPT$. 
    We consider a rounded solution profile $\roundsol$ with the same path $\vec{\hat{p}}=\vec{p}^*$, but rounded speed planning and charging planning. In particular, we set
    \begin{subequations}
        \begin{align}
            \hat{t}^r_{p_{i-1}, p_{i}} &= \hat{t}_{p_{i-1}, p_{i}}( \lceil c(t_{p_{i-1}, p_{i}}^{r*}) / \delta_\beta \rceil \delta_\beta,  \delta_\beta), \\
            \hat{\beta^c_j} &=  \lfloor \beta^{c*}_j / \delta_\beta \rfloor \delta_\beta,
        \end{align} 
    \end{subequations}
    where $\beta^{c*}_j, \hat{\beta}^c_j$ denotes the charged amount of energy at charging station $j$ of the optimal solution and the rounded solution respectively. We can then uniquely determine the charging time $\hat{t}^c_j$ given the such construction. After determining the travel time and the charging time, we can obtain the SoC profile $\vec{\hat{\beta}}$. 
    We then determine the waiting time $\hat{t}_w$ by solving problem~\eqref{prob:g:tw} for each charging station with the rounded objective, that is
    \begin{align}
        \hat{t}_j^w = g_j(\hat{\beta}_j, \hat{t}^c_j, \hat{\tau}_j, \lceil F_j^* / \delta_F \rceil \delta_F ) \label{lem:apxtest:tw}
    \end{align}
    where $F_j^*$ is the objective of the optimal solution incurred at the charging station $j$. 
    We note that by construction, the rounded solution always arrive earlier than the optimal solution at the charging station $j$. Therefore, $t^w = (t^{w*}_j + \tau^*_j - \hat{\tau}_j)$ is a feasible solution to the above problem~\eqref{lem:apxtest:tw} and the problem itself is feasible.

    We then show that such rounded solution $\roundsol$ satisfies the conditions in~\eqref{lem:apxtest:ineq}. By construction, we consume more energy on each road segment, wait less, and charge less at the charging station. Thus, each time decision of the rounded solution is no greater than its corresponding optimal time decision, and we have
    \begin{align}
        \sum_{i=1}^{n} \hat{t}^r_i + \sum_{i=1}^{m} (\hat{t}^w_i + \hat{t}^c_i) \leq  \sum_{i=1}^{n} t^{r*}_i + \sum_{i=1}^{m} (t^{w*}_i + t^{c*}_i) \leq T. 
    \end{align}
    Meanwhile, difference of the SoC evolution between optimal solution and and the rounded solution is bounded by the discretization precision $\delta_\beta$, i.e.,
    \begin{align}
        (\beta_{i+1}^* - \beta_{i}^*) - \delta_\beta \leq (\hat{\beta}_{i+1} - \hat{\beta}_{i}) \leq (\beta_{i+1}^* - \beta_{i}^*) 
    \end{align}
    Therefore, we have
    \begin{subequations}
        \begin{align}
            \hat{\beta}_i & = \sum_{j=0}^{i} (\hat{\beta}_{j+1} - \hat{\beta}_{j}) + \hat{\beta}_0 
                \\ & \geq \sum_{j=0}^i (\beta_{j+1}^* - \beta_{j}^* - \delta_\beta) + \hat{\beta}_0
                \\ &= \beta_i^* - i \delta_\beta + (\hat{\beta}_0 - \beta_0 )
                \\ &\geq \beta_i^* - (|V|+|V_c|)\delta_\beta + \epsilon_\beta B
                \\ &\geq 0
        \end{align}
    \end{subequations}
    For the other direction, we also have
    \begin{align}
        \hat{\beta}_i \leq \beta_i^* + (\hat{\beta}_0 - \beta_0) \leq (1+\epsilon_\beta) B.
    \end{align}
    Next, we show that the rounded solution satisfies the objective bound~\eqref{lem:apxtest:ineq:obj}. Because the problem~\eqref{lem:apxtest:tw} is always feasible, the incurred objective $\hat{F}_j$ of the rounded solution at the charging station $j$ satisfies 
    \begin{align}
        \hat{F}_j \leq \lceil F_j^* / \delta_F \rceil \delta_F \leq (F_j^*/\delta_F + 1) \delta_F = F_j^* + \delta_F.
    \end{align}
    Therefore, by summing up all the objectives at each charging station, we have
    \begin{subequations}
        \begin{align}
            F(\vec{\hat{p}}, \vec{\hat{t}}^r, \vec{\hat{t}}^w, \vec{\hat{t}}^{c}) & = \sum_{j=1}^{m} \hat{F}_j \leq \sum_{j=1}^{m} (F_j^* + \delta_F) 
            \\ & \leq \OPT + m\delta_F \leq \OPT + |V_c| \delta_F
            \\ & \leq \OPT + \epsilon_F \Omega.
        \end{align}
    \end{subequations}
    Meanwhile, by construction and the DP structure of Algorithm~\ref{alg:apxtest}, given the assumption that $\Omega \geq \OPT$, the rounded solution must be examined in the Algorithm~\ref{alg:apxtest}. Therefore, the algorithm must return in line~\ref{alg:apxtest:return} and the produced solution has no larger objective than the rounded solution~$\roundsol$. 
    Meanwhile, the produced solution naturally satisfies the conditions~\eqref{lem:apxtest:ineq:ddl} and~\eqref{lem:apxtest:ineq:b}. This completes our proof.

    
\end{proof}}

\end{document}